\documentstyle[prd,aps,psfig]{revtex}

 \newcommand{\CL}{{\cal L}}

 \newcommand{\bea}{\begin{eqnarray}}  \newcommand{\eea}{\end{eqnarray}}
 \newcommand{\beq}{\begin{equation}}  \newcommand{\eeq}{\end{equation}}
 \newcommand{\non}{\nonumber}  
   
 \newcommand{\lmk}{\left(}  \newcommand{\rmk}{\right)}

 \newcommand{\del}{\partial}  
 \newcommand{\vect}[1]{\mbox{\boldmath${#1}$}}
 \newcommand{\bib}{\bibitem}

\def\IB#1#2#3{{\bf #1}, #2 (19#3)}

\def\IBID#1#2#3{{\it ibid}. {\bf #1}, #2 (19#3)}
\def\IBIDD#1#2#3{{\it ibid}. {\bf #1}, #2 (20#3)}

\def\APJL#1#2#3{Astrophys. J. Lett. {\bf #1}, L#2 (19#3)}
\def\APJLL#1#2#3{Astrophys. J. Lett. {\bf #1}, L#2 (20#3)}

\def\IJMPD#1#2#3{Int. J. Mod. Phys. D {\bf #1}, #2 (19#3)}

\def\NATT#1#2#3{Nature (London) {\bf #1}, #2 (20#3)}

\def\PLB#1#2#3{Phys. Lett. B {\bf #1}, #2 (19#3)}

\def\PRD#1#2#3{Phys. Rev. D {\bf #1}, #2 (19#3)}
\def\PRDD#1#2#3{Phys. Rev. D {\bf #1}, #2 (20#3)}

\def\PRL#1#2#3{Phys. Rev. Lett. {\bf#1}, #2 (19#3)}

\def\PTP#1#2#3{Prog. Theor. Phys. {\bf #1}, #2 (19#3)}

\begin{document}
\draft
\title{Scaling property and peculiar velocity of global monopoles}
\author{Masahide Yamaguchi}
\address{Research Center for the Early Universe, University of Tokyo,
Tokyo, 113-0033, Japan}
\date{\today}
\maketitle

\begin{abstract} 
    We investigate the scaling property of global monopoles in the
    expanding universe. By directly solving the equations of motion
    for scalar fields, we follow the time development of the number
    density of global monopoles in the radiation dominated (RD)
    universe and the matter dominated (MD) universe. It is confirmed
    that the global monopole network relaxes into the scaling regime
    and the number per hubble volume is a constant irrespective of the
    cosmic time. The number density $n(t)$ of global monopoles is
    given by $n(t) \simeq (0.43\pm0.07) / t^{3}$ during the RD era and
    $n(t) \simeq (0.25\pm0.05) / t^{3}$ during the MD era. We also
    examine the peculiar velocity $v$ of global monopoles. For this
    purpose, we establish a method to measure the peculiar velocity by
    use of only the local quantities of the scalar fields. It is found
    that $v \sim (1.0 \pm 0.3)$ during the RD era and $v \sim (0.8 \pm
    0.3)$ during the MD era. By use of it, a more accurate analytic
    estimate for the number density of global monopoles is obtained.
\end{abstract}


\section{Introduction}

\label{sec:int}

The grand unified theory based on a simple group predicts magnetic
(gauge) monopoles if it breaks to leave the $U(1)$ symmetry of the
electromagnetism \cite{Preskill}. Magnetic monopoles are dangerous
because they may overclose our universe \cite{Preskill}. However,
no magnetic monopoles have been found yet. In fact, the flux of
magnetic monopoles is severely constrained by cosmological and
astrophysical considerations \cite{TD}. Thus, magnetic monopoles
produced in the early universe must be diluted away, annihilated, or
swept away (the monopole problem). This monopole problem is one of the
motivations of inflation \cite{inflation} though other solutions are
also proposed \cite{LP,DLV}.

On the other hand, global monopoles have drawn less attention.
However, while magnetic monopoles are dangerous for the cosmic
history, global monopoles may be favorable because they may produce
primordial density fluctuations responsible for the large scale
structure formation and the anisotropy of the cosmic microwave
background radiation (CMB) \cite{BV,BR1,BR2,PST,DKM}. Recent
observations of the CMB by the Boomerang \cite{BOOMERANG} and the
MAXIMA \cite{MAXIMA} experiments found the first acoustic peak with a
spherical harmonic multipole $l \sim 200$ predicted by the standard
inflationary scenario. But, they also found a relatively low second
peak, which may suggest the contribution of topological defects
\cite{TDCMB}. Moreover, deviations from Gaussianity in CMB are
reported in \cite{nonG}. Thus, though it is improbable for topological
defects to become the primary source of primordial density
fluctuations, a hybrid model is still attractive, where primordial
density fluctuations are comprised of adiabatic fluctuations induced
by inflation and isocurvature ones induced by topological defects. In
fact, topological defects can be easily compatible with inflation
\cite{hybrid,Yokoyama,preheating}.

The key property of global monopoles to contribute primordial density
fluctuations properly is scaling, where the typical scale of the
global monopole network grows in proportion to the horizon
scale.\footnote{For both the gauge \cite{local,VHS} and the global
string network \cite{global}, the scaling property is confirmed so
that density fluctuations produced by them become scale invariant.}
Then the number density of global monopoles is proportional to
$t^{-3}$ ($t$ : the cosmic time). Here we define the scaling parameter
$\xi$ as
\beq
   \xi \equiv n(t) t^3,
\eeq
where $n(t)$ is the number density of global monopoles. If $\xi$
becomes a constant irrespective of the cosmic time, we can conclude
that the global monopole network goes into the scaling regime.

The mass of a global monopole $m$ is proportional to the distance to
the nearest neighborhood antimonopole $d$ [$m \simeq 4 \pi \eta^{2}
d$, $\eta$ : the absolute magnitude of vacuum expectation values (VEV)
of scalar fields]. Roughly speaking, the distance is the horizon scale
(to be exact $d \simeq t / \xi^{1/3}$ ) if the network follows the
scaling property. Then the density fluctuations produced by global
monopoles are given by
\bea
  \frac{\delta\rho}{\rho} &\simeq& m n~8 \pi / 3 m_{\rm pl}^{2} H^{2} 
                                    \non \\
                          &\simeq& \frac{128}{3}~(24)~\pi^{2}
                                    \xi^{2/3} 
                                     \lmk \frac{\eta}{m_{\rm pl}}
                                      \rmk^{2}
                           \qquad {\rm for~RD~(MD)},
\eea
where $m_{\rm pl} = 1.2 \times 10^{19}$~GeV is the Plank mass. Thus,
the density fluctuations produced by global monopoles become scale
invariant. One may wonder if the precise value of $\xi$ is not so
important because the amplitude of density fluctuations depends on
only the combination of $\xi^{2/3} \eta^{2}$. It is true for the case
where global monopoles are the dominant source of density
fluctuations. But that is not the case. As stated earlier, the recent
observations indicate that the dominant source of density fluctuations
is inflation and topological defects contribute to them subdominantly.
In such a case, $\xi$ itself determines the ratio of the above two
contributions for a fixed VEV. Thus, the precise value of $\xi$ is
increasingly important.

The cosmological evolution of global monopoles was first discussed by
Barriola and Vilenkin \cite{BV}. They showed that the annihilation is
so efficient that global monopoles do not overclose our universe
unlike magnetic monopoles, but that it is not too efficient to survive
in our universe. This is mainly because an attractive force works
between a monopole and an antimonopole. Then, Bennett and Rhie
performed the first numerical simulations and found the tendency that
the number of global monopoles per horizon volume is nearly a constant
\cite{BR1}. However they used the nonlinear $\sigma$ model
approximation as equations of motion to evolve the scalar fields.
Later, Pen, Spergel, and Turok made numerical simulations in both the
nonlinear $\sigma$ model approximation and the full potential
\cite{PST}. (See also \cite{DKM}.) However, due to the lack of the
computer power, they can run a few realizations so that the scaling
property cannot be confirmed definitely. Many realizations of
numerical simulations are needed to decrease the error and estimate it
statistically.  Furthermore, in order to confirm the scaling property
completely, we should pay attention to several effects, which may
affect the final result, for example, the boundary effect, the grid
size effect, the total box size dependence, and so on.

In the previous paper \cite{Yamaguchi}, we reported the results of our
numerical simulations of the global monopole network and confirmed
that the global monopole network relaxes into the scaling regime both
in the RD and the MD universe. In this paper, we investigate the
cosmological evolution of the global monopole network comprehensively.

In the next section, we give the formulation and the results of our
numerical simulations. From the symmetry restoration phase, we follow
the evolution of the scalar fields with the $O(3)$ symmetry, which
breaks to generate global monopoles. The time development of the
number density of global monopoles is examined. Since we need to
perform a lot of realizations, it is important to establish the method
to identify monopoles automatically from the values of the scalar
fields.  We will propose two identification methods and compare the
results obtained by both methods. We also investigate the peculiar
velocity of global monopoles. Our numerical simulations are based on
the Eulerian view. Therefore, it is very difficult to know where a
monopole moves after the long interval enough to measure the velocity.
We establish the method to measure the velocity of global monopoles by
use of only the local quantities of scalar fields. In Sec. III, we set
up the Boltzmann equation for the time development of the number
density of global monopoles. Using the peculiar velocity obtained from
numerical simulations, an analytic estimate for the number density of
global monopoles is given. In the final section, we give the summary.

\section{Numerical Simulations}

\label{sec:num}

First of all, we give the formalism of our numerical simulations to
follow the evolution of the global monopole network. Later we show the
results of our numerical simulations and discuss whether the global
monopole network goes into the scaling regime. Furthermore, the
peculiar velocity of global monopoles is investigated.

We directly solve the equations of motion for scalar fields in the
expanding universe, which have the $O(3)$ symmetry at high temperature
and later break to generate global monopoles. We consider the
following Lagrangian density for scalar fields $\phi^{a}(x) (a = 1, 2,
3)$:
\beq
  \CL[\phi^{a}] = \frac12 g_{\mu\nu}\del^{\mu}\phi^{a}\del^{\nu}\phi^{a}
                 - V_{\rm eff}[\phi^{a},T].
\eeq 
Here $g_{\mu\nu}$ is the flat Robertson-Walker metric and the
effective potential $V_{\rm eff}[\phi^{a},T]$, which represents the
typical second order phase transition, is given by
\bea
  V_{\rm eff}[\phi^{a},T] 
       &=& \frac{\lambda}{4}(\phi^{2} - \sigma^2)^2 
                 + \frac{5}{24}\lambda T^2 \phi^{2}, \non \\
       &=& \frac{\lambda}{4}(\phi^{2} - \eta^2)^2 
            + \frac{\lambda}{4}(\sigma^{4} - \eta^4), 
  \label{eqn:effpot}
\eea
where $\phi \equiv \sqrt{\phi^{a}\phi^{a}}$, $\eta \equiv \sigma
\sqrt{1-(T/T_{c})^2}$ and $T_{c} \equiv \frac25\sqrt{15}\sigma$ is the
critical temperature. For $T > T_{c}$, the potential $V_{\rm
eff}[\phi^{a},T]$ has a minimum at the origin and the $O(3)$ symmetry
is restored. On the other hand, for $T < T_{c}$, new minima $\phi =
\eta$ appear and the symmetry is broken, which leads to the formation
of global monopoles.

The equations of motion for the scalar fields $\phi^{a}$ in the
expanding universe are given by
\beq
  \ddot{\phi^{a}}(x) + 3H\dot{\phi^{a}}(x) -
     \frac{1}{R(t)^2} \nabla^2 \phi^{a}(x)
   + \lambda \lmk \phi^{2}(x) - \eta^{2} \rmk \phi^{a}(x) 
    = 0,
  \label{eqn:master}
\eeq
where the dot represents the time derivative and $R(t)$ is the cosmic
scale factor. The Hubble parameter $H = \dot R(t)/R(t)$ and the cosmic
time $t$ are given by
\bea
  H^2 &=& \frac{4\pi^{3}}{45 m_{\rm pl}^2} g_{*} T^4,
   ~~~
  t = \frac{1}{2H} \equiv \frac{\epsilon_{RD}}{T^2} 
    ~~~~~~~~({\rm for~RD}), \non \\
  H^2 &=& \alpha(T) \frac{4\pi^{3}}{45 m_{\rm pl}^2} g_{*} T^4,
   ~~
  t = \frac{2}{3H} \equiv \frac{\epsilon_{MD}}{T^{3/2}} 
    ~~({\rm for~MD}),  
  \label{eqn:hubble}
\eea
with $g_{*}$ to be the total number of degrees of freedom for the
relativistic particles. For the MD case, we have defined $\alpha(T)$
[$\alpha(T) > 1$] as $\alpha(T) \equiv \rho_{\rm mat}(T) / \rho_{\rm
rad}(T) = \alpha_c (T_{c}/T)$, where $\rho_{\rm mat}(T)$ is the
contribution to the energy density from nonrelativistic particles,
$\rho_{\rm rad}(T)$ is the contribution from relativistic particles at
the temperature $T$, and $\alpha_{c} \equiv \rho_{\rm mat}(T_{c}) /
\rho_{\rm rad}(T_{c})$. We also define the dimensionless parameter
$\zeta$ as
\bea
  \zeta_{RD} &\equiv& \frac{\epsilon_{RD}}{\sigma}  =
     \lmk \frac{45}{16\pi^3g_{*}} \rmk^{1/2}
     \frac{m_{\rm pl}}{\sigma} 
    ~~~~~~~({\rm for~RD}),  \non \\
  \zeta_{MD} &\equiv& \frac{\epsilon_{MD}}{\sigma^{1/2}}  =
     \lmk \frac{5\sqrt{15}}{6\alpha_{c}\pi^3g_{*}} \rmk^{1/2}
     \frac{m_{\rm pl}}{\sigma}
    ~~~~({\rm for~MD}).  
  \label{eqn:zeta}
\eea
In our simulation, we take $\zeta_{RD,MD} = 10$ and $5$ to investigate
the $\zeta$ dependence on the result.

We start the numerical simulations from the $O(3)$ symmetric phase
with the temperature $T_{i} = 2T_{c}$, which corresponds to $t_{i} =
t_{c}/4$(RD) and $t_{i} = t_{c}/(2\sqrt{2})$(MD). At the initial time
($t_{i} < t_{c}$), we adopt as the initial condition the thermal
equilibrium state with the mass
\beq
  m = \sqrt{\frac{5}{12}\lambda(T_{i}^{2} - T_{c}^{2})},
\eeq
which is the inverse curvature of the potential at the origin at $t =
t_{i}$.

Hereafter we normalize the scalar field in units of $t_{i}^{-1}$, $t$
and $x$ in units of $t_{i}$. We set $\lambda$ to be $\lambda = 0.25$
and normalize the scale factor $R(t)$ as $R(1) = 1$. Then, the
normalized equations of motion for the scalar fields $\phi^{a}$ are
given by
\bea
  && \ddot{\phi^{a}}(x) + \frac{3}{2t}\dot{\phi^{a}}(x) -
    \frac{1}{t}\nabla^2\phi^{a}(x) 
      + \lambda (\phi^{2} - \eta_{RD}^{2}) \phi^{a} = 0
     \hspace{2cm} ({\rm for~RD}), \non \\
  && \ddot{\phi^{a}}(x) + \frac{2}{t}\dot{\phi^{a}}(x) -
    \frac{1}{t^{4/3}}\nabla^2\phi^{a}(x) 
      + \lambda (\phi^{2} - \eta_{MD}^{2}) \phi^{a} = 0
     \hspace{1.6cm} ({\rm for~MD})
\eea
with $\eta_{RD} = \frac{5}{48}\zeta_{RD}\sqrt{1-4/t}$ and $\eta_{MD} =
(\sqrt{15}/12)^{3/2}\zeta_{MD}\sqrt{1-4/t^{4/3}}$.
 
We perform numerical simulations in seven different sets of lattice
sizes and lattice spacings in the RD universe and the MD universe (See
Tables \ref{tab:set1} and \ref{tab:set2}.). In all cases, the time
step is taken to be $\delta t = 0.01$. In the typical case (1), the
box size is nearly equal to the horizon volume $(H^{-1})^{3}$ and the
lattice spacing to the typical core size of a monopole $\delta x \sim
1.0/(\sqrt{\lambda}\sigma)$ at the final time $t_{f}$. Furthermore, in
order to investigate the dependence of $\zeta$, we arrange the case
(7) with $\zeta = 5$. We have simulated the system from 10~[(2), (3),
(5) and (6)] or 50~[(1), (4), and (7)] different thermal initial
conditions. Also, in order to investigate the effect of the boundary
condition (BC), we adopt both the periodic BC and the reflective BC
[$\nabla^{2}\phi^{a}(x) = 0$ on the boundary].

\subsection{Number density}

In order to judge whether the global monopole network relaxes into the
scaling regime, we follow the time development of $\xi$. If $\xi$
becomes a constant irrespective of the cosmic time, we can conclude
that the global monopole network goes into the scaling regime.

First of all, we need to count the number of global monopoles in the
simulation box. For the purpose, we must establish the identification
method of global monopoles because we can obtain only the values of
scalar fields. We propose two identification methods and later compare
the results obtained by using both methods. In the first method (I),
we use a static spherically-symmetric solution with the topological
charge $N = 1$, which is obtained by solving the equation
\beq
  \frac{d^2 \phi}{dr^2} + \frac{2}{r}\frac{d\phi}{dr}
   - 2\frac{\phi}{r^2} - \frac{dV_{\rm eff}[\phi,T]}{d\phi} = 0
\eeq
with $\phi^{a}(r, \theta, \varphi) \equiv \phi(r)x^{a} / r$, $x^{1} =
r\sin\theta\cos\varphi, x^{2} = r\sin\theta\sin\varphi$, and $x^{3} =
r\cos\theta$. The boundary conditions are given by
\bea
  \phi(r) &\rightarrow& \eta \qquad (r \rightarrow \infty), 
     \non \\
  \phi(0) &=& 0.
\eea
One should notice that a point with $\phi^{a} = 0$ for all $a$'s is
not necessarily situated at a lattice point. In the worst case, a
point with $\phi^{a} = 0$ lies at the center of a cube. Then, we make
the criterion that a lattice is identified with a part of a monopole
core if the potential energy density there is larger than that
corresponding to the field value of a static spherically-symmetric
solution at $r = \sqrt{3}\delta x_{\rm{phys}}/2$ [$\delta
x_{\rm{phys}} = R(t)\delta x$], that is, the potential energy density
at the vertices when a static spherically-symmetric monopole lies at
the center of cube.  Moreover, in order to reduce the error, we look
on the identified lattices which are connected as one monopole core.
In the other method (II), a cubic box is regarded as including a
monopole if all $\phi^{a} = 0$ ($a$ = 1, 2, 3) surfaces pass through
the cubic box, that is, the signs of all eight vertices of the box are
not identical for each three field $\phi^{a}$. In this method, we also
look on the identified boxes which are connected as one monopole core.
In fact, as shown later, the results with these two identification
methods coincide very well.

\begin{figure}
  \begin{center}
    \leavevmode\psfig{figure=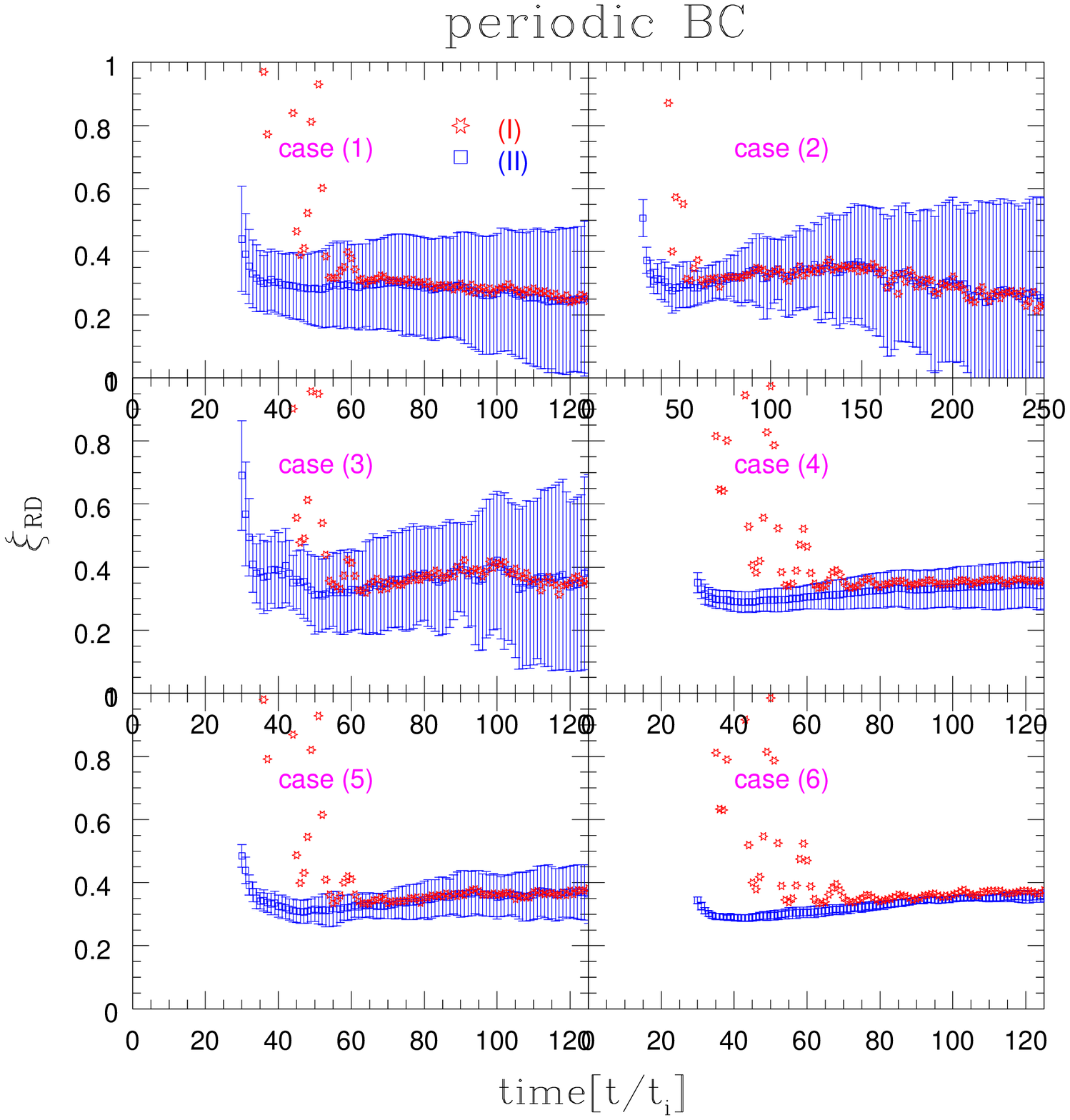,width=19.0cm}
  \end{center}
  \caption{The time development of $\xi_{RD}$ in the cases from (1) to
  (6) under the periodic BC for the RD case. Asterisks ($\ast$)
  represent time development of $\xi_{RD}$ for the identification
  method (I). Squares ($\Box$) represent time development of
  $\xi_{RD}$ for the identification method (II). The vertical lines
  denote a standard deviation over different initial conditions for
  the identification method (II).}
  \label{fig:xiRDp}
\end{figure}

\begin{figure}
  \begin{center}
    \leavevmode\psfig{figure=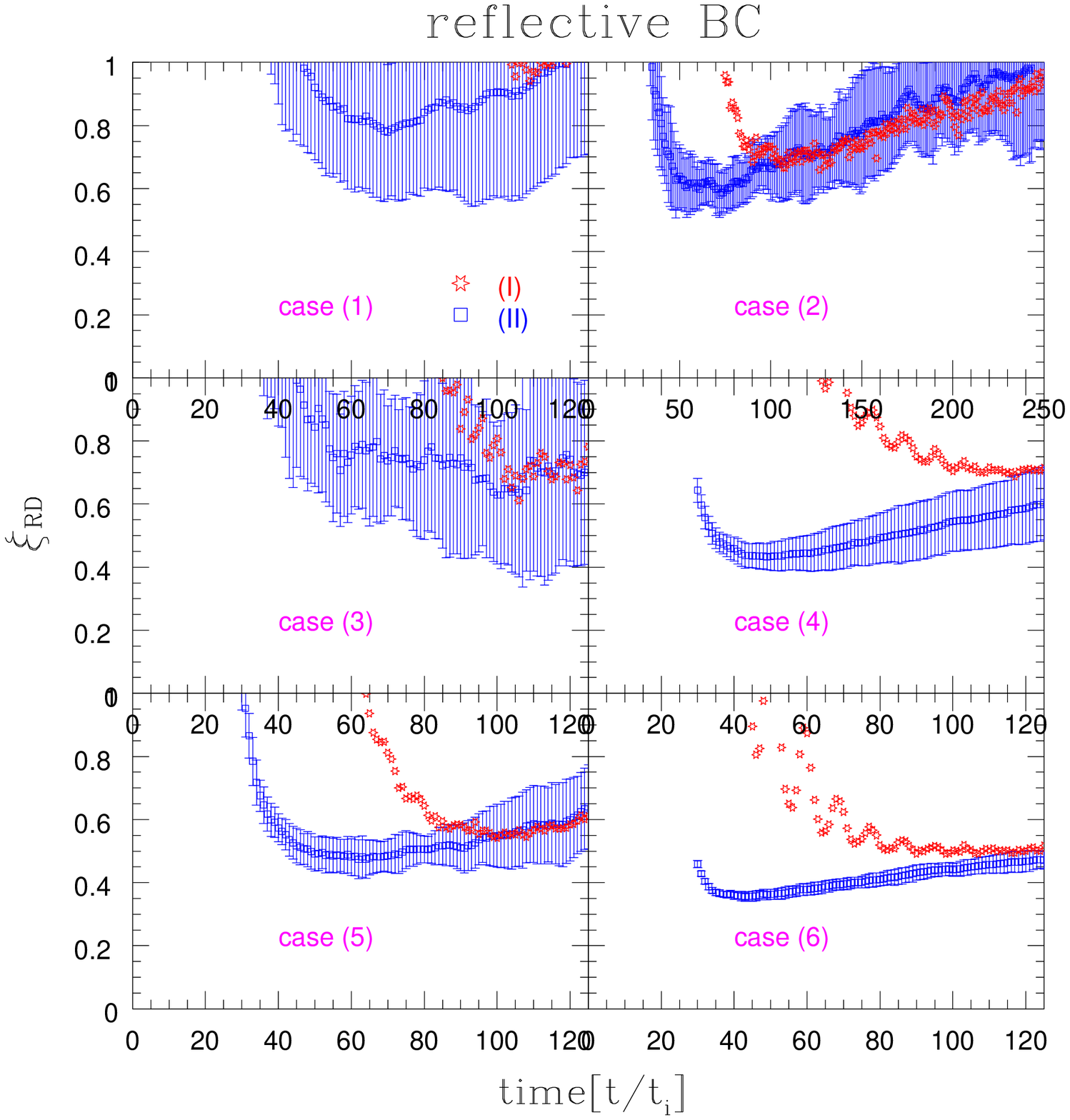,width=19.0cm}
  \end{center}
  \caption{That under the reflective BC.}
  \label{fig:xiRDr}
\end{figure}

\begin{figure}
  \begin{center}
    \leavevmode\psfig{figure=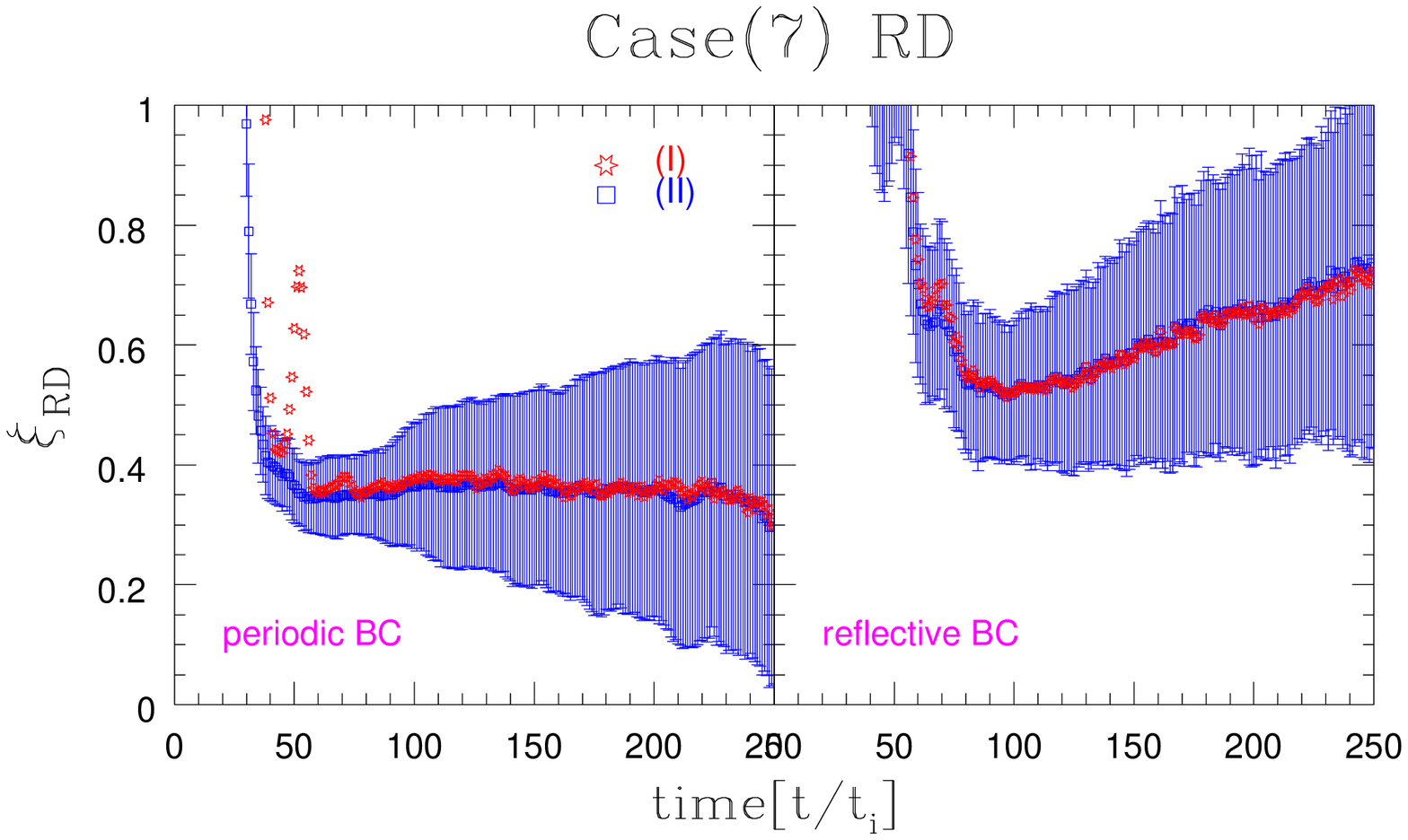,width=19.0cm}
  \end{center}
  \caption{The time development of $\xi_{RD}$ in the RD universe for the
  case (7) under the periodic BC and the reflective BC.}
  \label{fig:xiRD11}
\end{figure}

First of all, we discuss the evolution of global monopoles in the RD
universe. The time development of $\xi_{RD}$ in the cases from (1) to
(6) under the periodic BC is described in Figs. \ref{fig:xiRDp}.
Asterisks ($\ast$) represent the time development of $\xi_{RD}$ for
the identification method (I). Squares ($\Box$) represent the time
development of $\xi_{RD}$ for the identification method (II). As
easily seen, the results with two identification methods coincide very
well. We also find that after some relaxation period, $\xi_{RD}$
becomes a constant irrespective of time for all cases. Though all
results are consistent within the standard deviation, $\xi_{RD}$ tends
to increase as the box size does. This is understood as follows: under
the periodic BC, a monopole can annihilate with an antimonopole which
lies beyond the boundary. Therefore monopoles under the periodic BC
annihilate more often than those in the real universe, in particular,
monopoles annihilate more often for smaller box sizes. On the other
hand, the time development of $\xi_{RD}$ in the cases from (1) to (6)
under the reflective BC is described in Figs. \ref{fig:xiRDr}. We also
find that, $\xi_{RD}$ tends to become a constant though more
relaxation period takes. Contrary to the case under the periodic BC,
$\xi_{RD}$ tends to decrease as the box size increases. This is
understood as follows: under the reflective BC, field configurations
of a monopole extend with the same phase direction beyond the boundary
so that a monopole cannot annihilate any antimonopoles which may lie
beyond the boundary in the real universe. Therefore monopoles under
the reflective BC annihilate less often than those in the real
universe, in particular, monopoles annihilate less often for smaller
box sizes. Thus, $\xi_{RD}$ takes a larger value in a smaller-box
simulation due to the boundary effect.\footnote{One may wonder if
$\xi_{RD}$ increases even after some relaxation period, particular, in
the case (2), which is the longest simulation. This is also just the
boundary effect because the earlier the cosmic time is, the simulation
box is larger than the horizon volume.} After all, the real number of
the monopole per the horizon volume lies in between those under the
periodic BC and the reflective BC. $\xi_{RD}$ of each case is listed
in Table \ref{tab:set1}. From the results of the largest-box
simulations [case (6)], we can conclude that the global monopole
network relaxes into scaling regime in the RD universe and $\xi_{RD}$
converges to a constant $\xi_{RD} \simeq (0.43 \pm 0.07)$. We also
show the time development of $\xi_{RD}$ with $\zeta = 5$ in the case
(7) (Fig. \ref{fig:xiRD11}). $\xi_{RD}$ asymptotically becomes a
constant $\xi_{RD} \simeq (0.36 \pm 0.01)$ under the periodic BC,
which is consistent with the above all cases with $\zeta = 10$ within
the standard deviation.\footnote{In this case, $\xi_{RD}$ under the
reflective BC also tends to increase due to the boundary effect.}
Hence we can also conclude that $\zeta$ does not change the essential
result.

\begin{figure}
  \begin{center}
    \leavevmode\psfig{figure=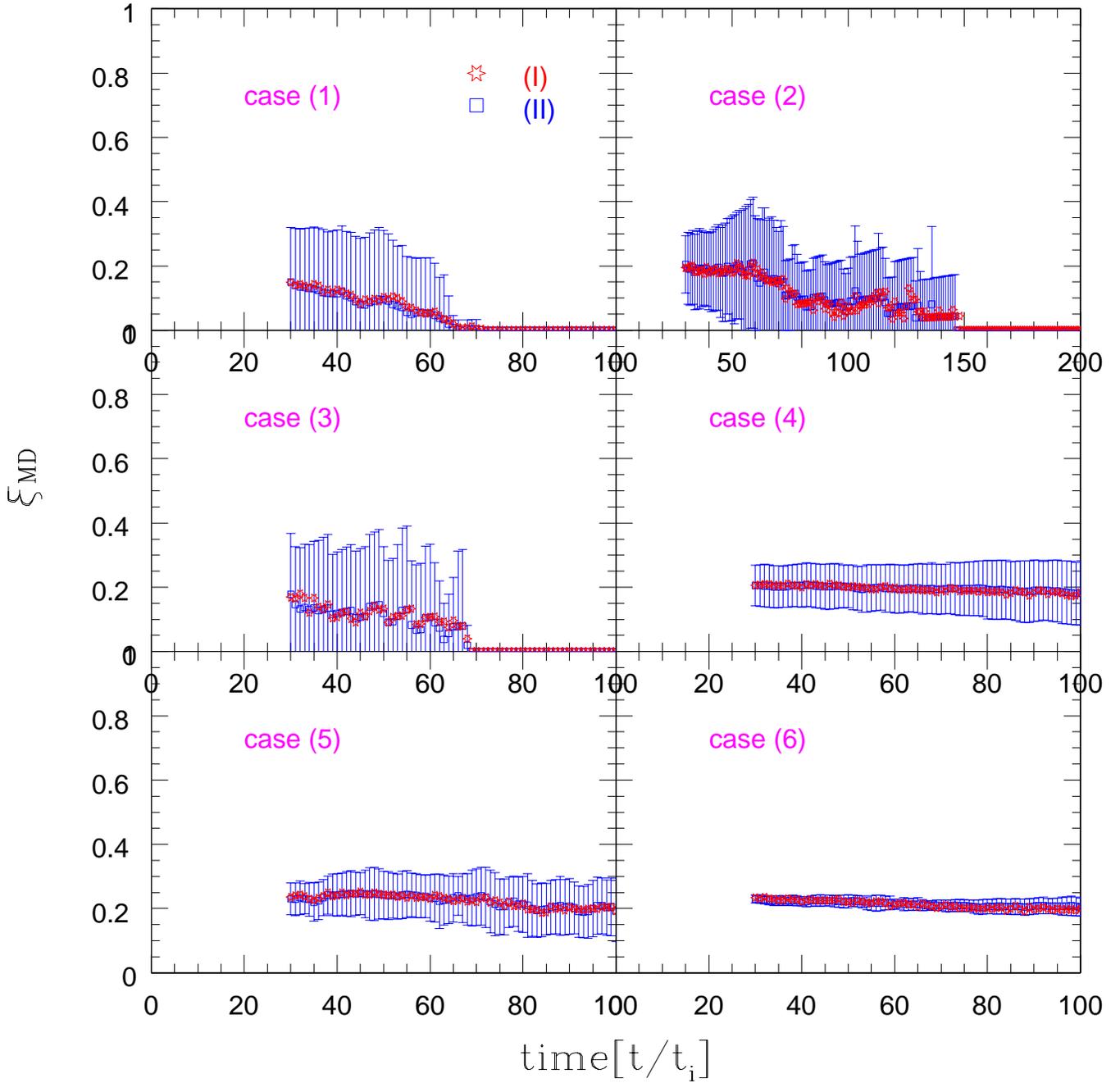,width=19.0cm}
  \end{center}
  \caption{The time development of $\xi_{MD}$ in the cases from (1) to
  (6) under the periodic BC for the MD case. Asterisks ($\ast$)
  represent time development of $\xi_{MD}$ for the identification
  method (I). Squares ($\Box$) represent time development of
  $\xi_{MD}$ for the identification method (II). The vertical lines
  denote a standard deviation over different initial conditions.}
  \label{fig:xiMDp}
\end{figure}

\begin{figure}
  \begin{center}
    \leavevmode\psfig{figure=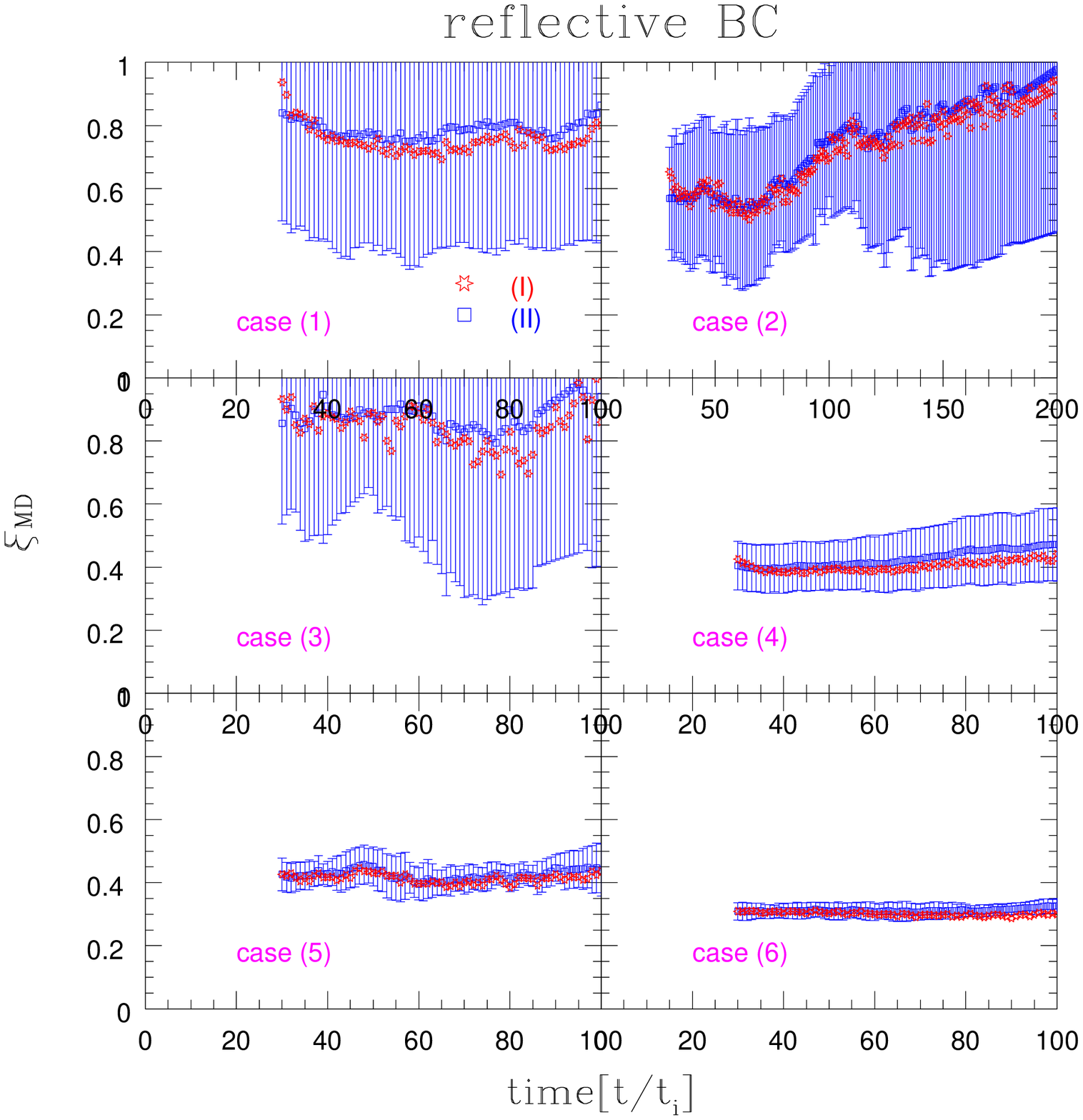,width=19.0cm}
  \end{center}
  \caption{That under the reflective BC.}
  \label{fig:xiMDr}
\end{figure}

\begin{figure}
  \begin{center}
    \leavevmode\psfig{figure=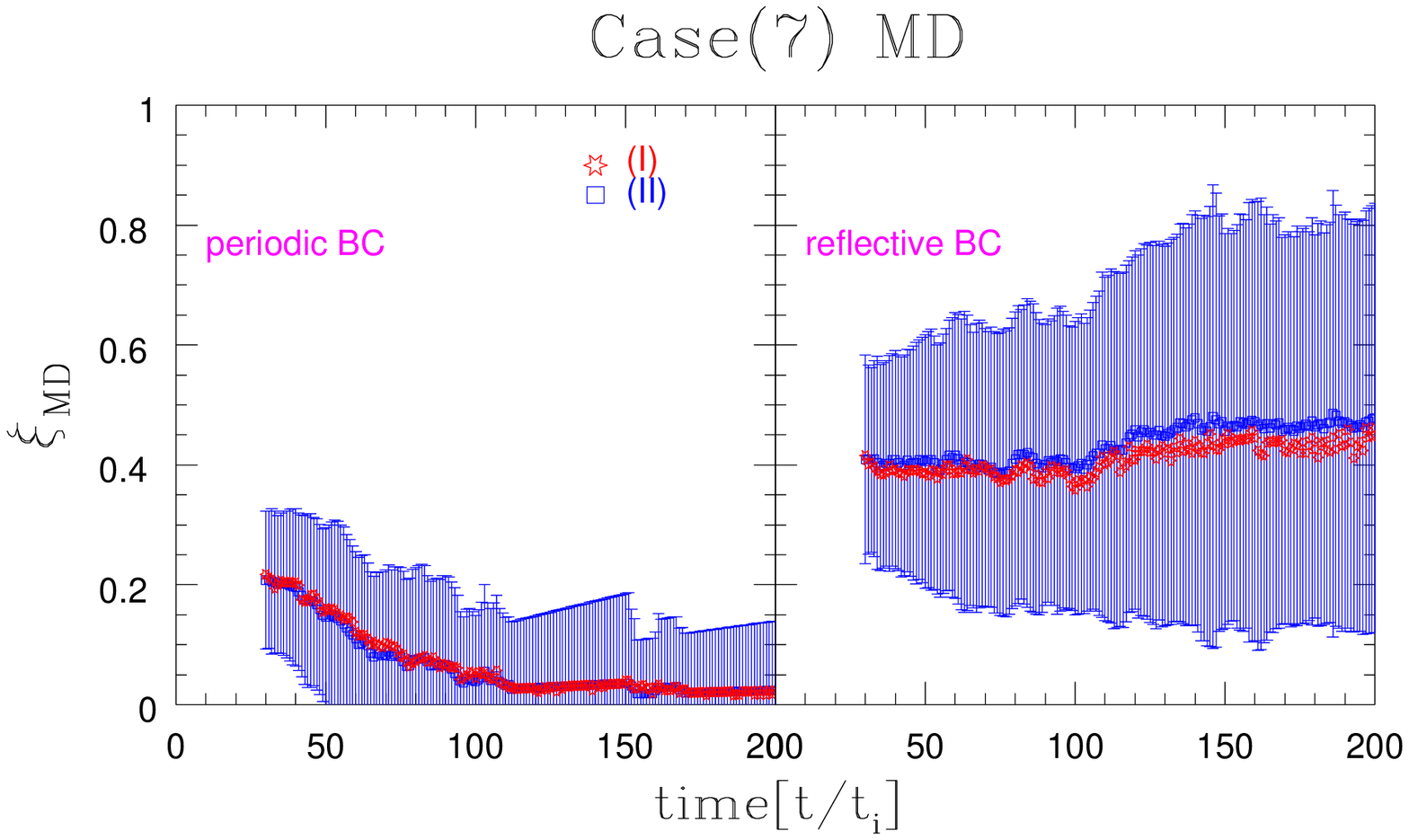,width=19.0cm}
  \end{center}
  \caption{The time development of $\xi_{MD}$ in the MD universe for the
  case (7) under the periodic BC and the reflective BC.}
  \label{fig:xiMD11}
\end{figure}

For the MD case, we also find that after some relaxation period, the
number of global monopoles per the horizon volume becomes a constant
irrespective of the cosmic time under the periodic BC except for the
cases (1), (2), and (3), in which global monopoles annihilate too much
due to the boundary effect. Also, the number of global monopoles per
the horizon volume becomes a constant irrespective of the cosmic time
under the reflective BC except for the case (2), where the boundary
effect is the most manifest because of the longest time simulation.
The tendency of the boundary effect is the same with that for the RD
case.  Then, from the results of the largest-box simulations [case
(6)], we conclude that the global monopole network relaxes into
scaling regime in the MD universe and $\xi_{MD}$ converges to a
constant $\xi_{MD} \simeq (0.25 \pm 0.05)$ (see Fig. \ref{fig:xiMDp}
and \ref{fig:xiMDr}). We also show the time development of $\xi_{MD}$
with $\zeta = 5$ in the case (7) (Fig.  \ref{fig:xiMD11}).  $\xi_{MD}$
asymptotically becomes a constant $\xi_{MD} \simeq (0.44 \pm 0.03)$
under the reflective BC though monopoles tend to disappear under the
periodic BC due to the boundary effect. This is consistent with the
above all cases with $\zeta = 10$ within the standard deviation. Thus,
we have completely confirmed that the global monopole network goes
into the scaling regime in both the RD universe and MD universe.

\subsection{Peculiar velocity}

In this subsection, we investigate the peculiar velocity of global
monopoles in the expanding universe. In order to measure the peculiar
velocity, we need to know where a monopole moves at the next step. For
the purpose, we need to find the method to look on the monopole found
at each time as the same. But, generally speaking, it is very
difficult in case there are a lot of monopoles in the simulation box.

Then, looking at the matter from another angle, we make best use of
the information of scalar fields. Since the values and the time
derivatives of scalar fields include all informations about monopoles,
the velocity of a monopole can be represented by only the information
of scalar fields, that is, local quantities. In fact, it is possible
as shown below. First of all, we expand scalar fields $\phi^{a}(\vect
x,t)$ around $\phi^{a}(\vect x_{0},t_{0})$ up to the first order,
\beq
  \phi^{a}(\vect x,t) \simeq \phi^{a}(\vect x_{0},t_{0}) 
      + \nabla \phi^{a}(\vect x_{0},t_{0}) \cdot (\vect x-\vect x_{0})
      + \dot\phi^{a}(\vect x_{0},t_{0}) (t-t_{0})
      \qquad ( a = 1, 2, 3 ).
\eeq
The monopole core is identified with the zero of all scalar fields
$\phi^{a}$. Assuming a monopole lies at $\vect x_{0}$ at the time
$t_{0}$, the position $\vect x$ of the monopole core at the
sufficiently near time $t$ is obtained as the intersection of the
following three planes,
\beq
  \vect A^{a} \cdot (\vect x-\vect x_{0}) + B^{a} (t-t_{0}) = 0,
\eeq
where $\vect A^{a} \equiv \nabla \phi^{a}(\vect x_{0},t_{0})$ and
$B^{a} \equiv \dot\phi^{a}(\vect x_{0},t_{0})$. These equations are
easily solved by the Cramer's formula
\beq
  \frac{(\vect x-\vect x_{0})_{j}}{t-t_{o}}
    = - \frac{ \begin{array}{ccc}
                                  &   j   &           \\
                 \vline A^{1}_{x} & B^{1} & A^{1}_{z} \vline \\
                 \vline A^{2}_{x} & B^{2} & A^{2}_{z} \vline \\
                 \vline A^{3}_{x} & B^{3} & A^{3}_{z} \vline
               \end{array} }
             { \begin{array}{ccc}
                 \vline A^{1}_{x} & A^{1}_{y} & A^{1}_{z} \vline \\
                 \vline A^{2}_{x} & A^{2}_{y} & A^{2}_{z} \vline \\
                 \vline A^{3}_{x} & A^{3}_{y} & A^{3}_{z} \vline
               \end{array} } ~.
\eeq
Thus, the peculiar velocity of a global monopole $v$ can be estimated
as \footnote{This method to measure the velocity of monopoles can
apply to that of strings in the same way. In the future publication,
we will investigate the velocity of the string network with the aid of
this method.}
\beq
  v = \frac{|\vect x-\vect x_{0}|}{t-t_{o}}.
\eeq

This method has two main sources to generate errors. First of all, our
estimate is correct only up to the first order. Then, in order to
reduce the error due to this approximation, we evaluate the peculiar
velocity for the case (3) in all the situations because it is the
simulation with the highest resolution (the smallest lattice spacing).
Next, a monopole does not necessarily lie just on the lattice in our
simulations. Especially, in the identification method (I), we have
identified a lattice with a part of the monopole core if the potential
energy at the lattice is larger than a critical value. Since we have
started the simulation from the thermal equilibrium states, there are
still small thermal fluctuations at late times, which may accidentally
lead to the large potential energy for the lattices not corresponding
to the monopole core. Thus, in the identification method (I), some
lattices which have nothing to do with the monopole core may be
identified with monopole cores. At such lattices, the velocities
obtained by the above formula may become extraordinarily large, which
causes the large error.  Then, in order to reduce the errors, we
introduce the cutoff for the velocity and abandon the velocities which
are larger than a cutoff at estimating the average and dispersion of
the velocity. We have chosen several cutoff values and investigated
their effects on the results.  We found that the results do not change
much if the cutoff is smaller than a value (of course, it need to be
larger than unity). Though the velocity is at most unity, we set the
cutoff to be 1.5 in order to use as many data as possible and reduce
the artificial effect.

\begin{figure}
  \begin{center}
    \leavevmode\psfig{figure=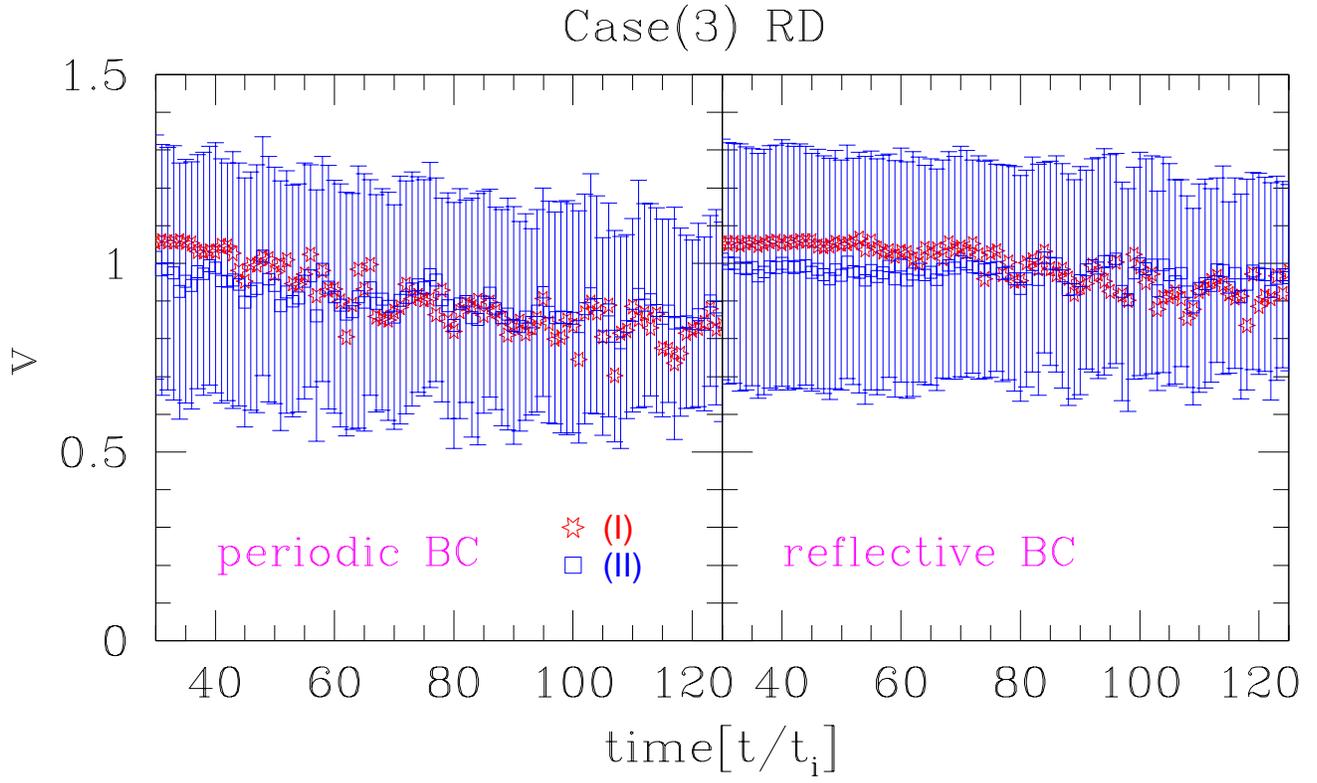,width=19.0cm}
  \end{center}
  \caption{The time development of the peculiar velocity in the RD
  universe for the case (3) under the periodic and the reflective BC.
  Asterisks ($\ast$) represent time development of the peculiar
  velocity $v$ for the identification method (I). Squares ($\Box$)
  represent time development of the peculiar velocity $v$ for the
  identification method (II). The vertical lines denote a standard
  deviation over different initial conditions for the identification
  method (II).}
  \label{fig:velocityr}
\end{figure}

\begin{figure}
  \begin{center}
    \leavevmode\psfig{figure=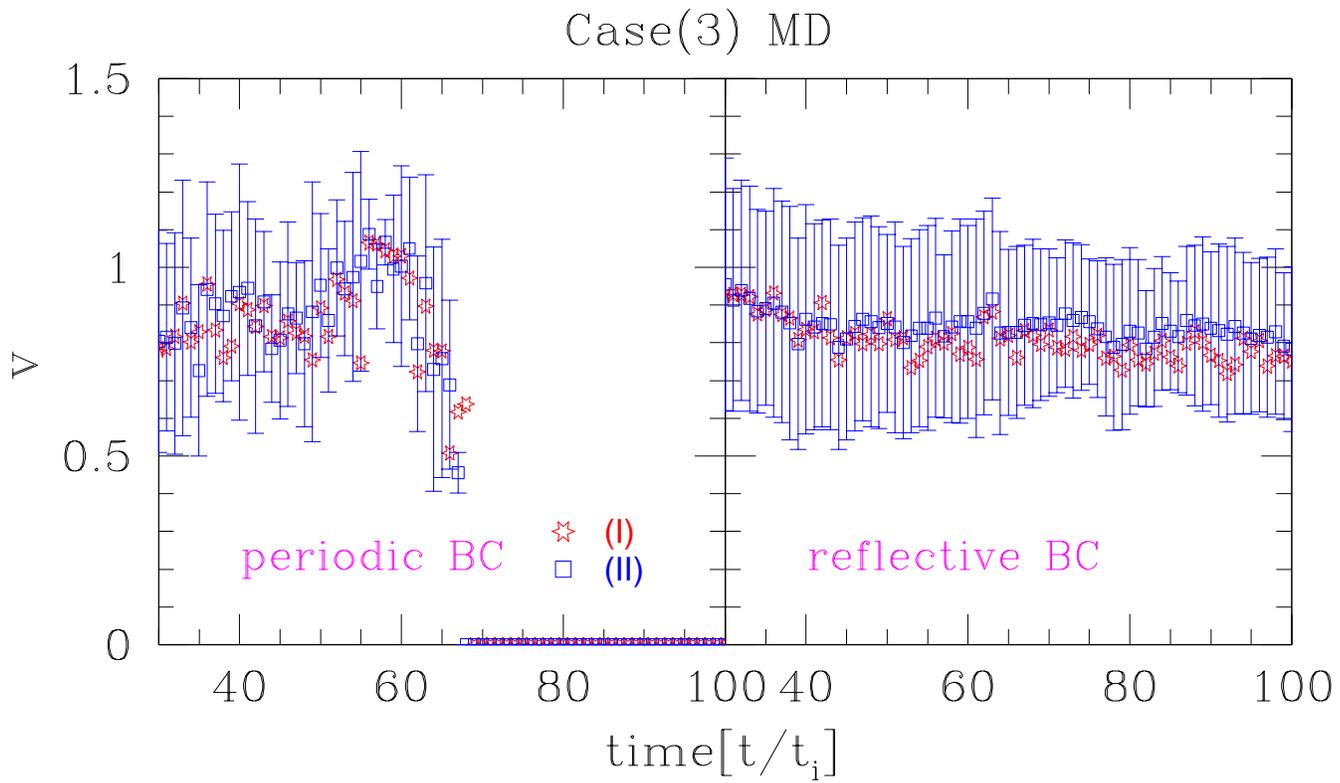,width=19.0cm}
  \end{center}
  \caption{The time development of the peculiar velocity in the MD
  universe for the case (3) under the periodic and the reflective BC.
  Since monopoles disappear at late times under the periodic BC, the
  peculiar velocity is set to zero in such a situation.}
  \label{fig:velocitym}
\end{figure}

The time development of the peculiar velocity of global monopoles
during the RD era under both the periodic and the reflective BCs is
depicted in Fig. \ref{fig:velocityr}. Though there is still large
uncertainty, the peculiar velocity $v$ takes almost the same constant
asymptotically under both BCs and is given by $v_{RD} \sim (1.0 \pm
0.3)$. On the other hand, the time development of the peculiar
velocity of global monopoles during the MD era is depicted in Fig.
\ref{fig:velocitym} and the peculiar velocity $v$ is given by $v_{MD}
\sim (0.8 \pm 0.3)$. Since monopoles disappear at late times under the
periodic BC, the peculiar velocity is set to zero in such a situation.

The obtained values of the peculiar velocity are roughly understood as
follows. A constant long-range attractive force works between a
monopole and an antimonopole due to the gradient energy of the scalar
fields. Then, the monopole is accelerated very much and the relative
velocity rapidly gets to the order of unity. In the RD universe, the
cosmic expansion is not so rapid that the redshift of the velocity due
to the cosmic expansion becomes negligible and the velocity reaches
almost the unity. On the other hand, in the MD era, the universe
expands so rapid that the velocity is redshifted and it takes a value
smaller than the unity.

\section{Analytic Estimate}

\label{sec:ana}

In this section, we give a simple analytic estimate for the scaling
parameter $\xi$.

The evolution for the number density of global monopoles $n(t)$ can be
described by the following Boltzmann equation,\footnote{A similar
discussion was done in \cite{Preskill,YYK}.}
\bea
  \frac{dn(t)}{dt} &=& - P(t) n(t) - 3 H(t) n(t), \non \\
                   &=& - \frac{n(t)}{T(t)} - \frac{3 m n(t)}{t},
  \label{eq:Bol}
\eea
where $R(t) \propto t^{m}$, $P(t)$ is the probability per unit time
that a monopole annihilates with an antimonopole, and $T(t)$ is the
period it takes for a pair of monopoles at rest with the mean
separation $l(t)$ to pair annihilate. The mean separation $l(t)$ is
given by $l(t) \equiv R(t) r_{s} = n(t)^{-1/3}$, where $r_{s}$ is the
mean comoving separation. In the previous publication
\cite{Yamaguchi}, we assumed that the relative velocity between them
reaches the order of unity at once because a constant attractive force
works between a pair of monopoles irrespective of the separation
length. In the previous section, we have confirmed that the above
assumption is basically correct but the peculiar velocity is smaller
than unity in the MD universe. Then, assuming that the relative
velocity is given by the peculiar velocity obtained in the previous
section and a pair of monopoles does not spiral around each other for
a long time, we give a more accurate analytic estimate for the number
density of global monopoles. The period $T(t)$ is given by the
following relation,
\beq
  v \int^{T+t_{0}}_{t_{0}} \frac{dt}{R(t)} \simeq \int^{r_{s}}_{0} dr,
\eeq
where $t_{0}$ is the initial time where a pair of monopoles are at
rest. Then, the period $T(t)$ reads
\beq
  T(t) \simeq \lmk \frac{1-m} {v\,t^{m} n(t)^{1/3}}
               \rmk^{\frac{1}{1-m}} \qquad\qquad 
     ({\rm for}~~t_{0} \ll T).
\eeq
Inserting this result into the Boltzmann equation (\ref{eq:Bol}), the
number density $n(t)$ takes the following asymptotic value:
\beq
  n(t) \simeq \frac{\,3^{3\,(1-m)}~(1-m)^{\,3\,(2-m)}}{v^{3} t^{3}} 
          \propto t^{-3}.
\eeq
From the above asymptotic form, first of all, we find that the number
density $n(t)$ is proportional to the inverse of the cosmic time
cubed, $t^{-3}$, which implies that $\xi$ becomes a constant
irrespective of the cosmic time. $\xi$ is also estimated as
\beq
  \xi = \frac{3^{\,3\,(1-m)}~(1-m)^{\,3\,(2-m)}}{v^{3}}.
\eeq
Inserting $m = 1/2$ and $v_{RD} \sim (1.0 \pm 0.3)$, $\xi_{RD} \sim
(0.45 \pm 0.22)$. On the other hand, inserting $m = 2/3$ and $v_{MD}
\sim (0.8 \pm 0.3)$, $\xi_{MD} \sim (0.17 \pm 0.13)$. Thus, $\xi$
obtained by the analytic estimates can well reproduce that obtained
from the numerical simulations both in the RD universe and the MD
universe.

\section{Summary}

\label{sec:con}

In this paper, we have discussed the evolution of the global monopole
network in the expanding universe. We have completely confirmed that
the global monopole network relaxes into the scaling regime, where the
number of global monopoles per the horizon volume is a constant. The
scaling parameter $\xi$ is given by $\xi_{RD} \simeq (0.43 \pm 0.07)$
in the RD universe and $\xi_{MD} \simeq (0.25 \pm 0.05)$ in the MD
universe. We also investigated the peculiar velocity of global
monopoles. First of all, we established the method to measure the
peculiar velocity by using only the local quantities of the scalar
fields. This method compensates the weak point of the Eulerian view
which our numerical simulations are based on, that is, we cannot
follow the motion of each monopole in detail. We find that the
peculiar velocity also becomes a constant irrespective of the cosmic
time and is given by $v_{RD} \sim (1.0 \pm 0.3)$ and $v_{MD} \sim (0.8
\pm 0.3)$ though there is still large uncertainty. By use of the
Boltzmann equation for the time development of the number density and
the peculiar velocity obtained from numerical simulations, we give a
simple analytic estimate for the number density, which can well
reproduce the results from the numerical simulations up to the
proportional coefficient $\xi$.

\subsection*{ACKNOWLEDGMENTS}

The author is grateful to J. Yokoyama for useful comments.  This work
was partially supported by the Japanese Grant-in-Aid for Scientific
Research from the Ministry of Education, Culture, Sports, Science, and 
Technology.

\newpage

\newpage

\newpage

\newpage

\newpage

\newpage

\newpage

\newpage


\begin{table}
\caption{Seven different sets of the simulations for the RD case.}
\label{tab:set1}
  \begin{center}
     \begin{tabular}{cccccccc}
         Case & Lattice &
         Lattice spacing ($\delta x$) & $\zeta$ & Realization 
         & ${\rm Box~size}/H^{-1}$
         & $\xi$  & $\xi$ \\
         & number & [unit = $t_{i}R(t)$] &  &  & (at final time) &
         (periodic B.C.) & (reflective B.C.) \\
        \hline
        (1) & $128^3$ & $\sqrt{3}/10$ & 10 & 50 & 1(at 125) &
        $0.28\pm0.19$ & $1.00\pm035$\\ 
        (2) & $256^3$ & $\sqrt{6}/20$ & 10 & 10 & 1(at 250) &
        $0.31\pm0.20$ & $0.83\pm0.15$\\
        (3) & $256^3$ & $\sqrt{3}/20$ & 10 & 10 & 1(at 125) &
        $0.35\pm0.21$ & $0.71\pm0.33$\\
        (4) & $128^3$ & $\sqrt{3}/5$  & 10 & 50 & 2(at 125) &
        $0.37\pm0.06$ & $0.71\pm0.11$\\
        (5) & $256^3$ & $\sqrt{3}/10$ & 10 & 10 & 2(at 125) &
        $0.36\pm0.07$ & $0.58\pm0.12$\\
        (6) & $256^3$ & $\sqrt{3}/5$  & 10 & 10 & 4(at 125) &
        $0.36\pm0.01$ & $0.50\pm0.03$\\
        \hline
        (7) & $128^3$ & $\sqrt{6}/10$ & 5  & 50 & 1(at 250) &
        $0.36\pm0.17$ & $0.61\pm0.21$\\
     \end{tabular}
  \end{center}
\end{table}
\begin{table}
\caption{Seven different sets of the simulations for the MD case.}
\label{tab:set2}
  \begin{center}
     \begin{tabular}{cccccccc}
         Case & Lattice &
         Lattice spacing ($\delta x$) & $\zeta$ & Realization 
         & ${\rm Box~size}/H^{-1}$
         & $\xi$  & $\xi$ \\
         & number & [unit = $t_{i}R(t)$] &  &  & (at final time) &
         (periodic B.C.) & (reflective B.C.) \\
        \hline
        (1) & $128^3$ & $3(100)^{1/3}/256$ & 10 & 50 & 1(at 100) &
        Disappearance & $0.75\pm0.36$\\ 
        (2) & $256^3$ & $3(200)^{1/3}/128$ & 10 & 10 & 1(at 200) &
        Disappearance & $0.77\pm0.10$\\
        (3) & $256^3$ & $3(100)^{1/3}/512$ & 10 & 10 & 1(at 100) &
        Disappearance & $0.82\pm0.50$\\
        (4) & $128^3$ & $3(100)^{1/3}/128$ & 10 & 50 & 2(at 100) &
        $0.19\pm0.09$ & $0.42\pm0.10$\\
        (5) & $256^3$ & $3(100)^{1/3}/256$ & 10 & 10 & 2(at 100) &
        $0.21\pm0.09$ & $0.41\pm0.05$\\
        (6) & $256^3$ & $3(100)^{1/3}/128$ & 10 & 10 & 4(at 100) &
        $0.20\pm0.02$ & $0.30\pm0.02$\\
        \hline
        (7) & $128^3$ & $3(200)^{1/3}/256$ & 5  & 50 & 1(at 200) &
        Disappearance & $0.44\pm0.03$ \\
     \end{tabular}
  \end{center}
\end{table}


\end{document}